\documentclass[a4paper,11pt]{article}
\usepackage{amssymb,amsmath,bm}
\usepackage{graphics,graphicx}
\usepackage{array,booktabs}
\usepackage{authblk}
\usepackage{slashed}
\usepackage{subfig}
\usepackage{cite}
\usepackage[margin=2.5cm]{geometry}
\usepackage[table,dvipsnames]{xcolor}
\definecolor{Dblue}{rgb}{0.1, 0.1, 0.8}
\definecolor{Lblue}{rgb}{0.22,0.51,0.9}
\definecolor{green}{RGB}{53,170,102}
\definecolor{Green}{rgb}{0.0, 0.5, 0.0}
\definecolor{Bred}{rgb}{0.8, 0.25, 0.33}
\usepackage{tikz}
\usepackage{multirow}
\usepackage{tikz-feynman}
\usepackage[linktocpage,
colorlinks = true,
linkcolor = Bred,
urlcolor  = Dblue,
citecolor = green,
anchorcolor = blue]{hyperref}
\usepackage{epsfig}
\numberwithin{equation}{section}
\usepackage{multirow}
\usepackage{cite}
\usepackage{xcolor}
\usepackage[margin=2.5cm]{geometry}
\usepackage{orcidlink}
\usepackage[utf8]{inputenc}
\usepackage{orcidlink}

\DeclareUnicodeCharacter{2212}{-}

\numberwithin{equation}{section}

\usepackage{epsfig}

\title{\textbf{Scale-Invariant Three-Component Dark Matter and Gravitational Wave Signals}}
\author[a]{Mojtaba Hosseini,\thanks{\href{mailto:mojtaba\textunderscore hosseini@semnan.ac.ir}{mojtaba\textunderscore hosseini@semnan.ac.ir}}\orcidlink{0009-0000-4711-2083}}
\author[b]{Rouzbeh Rouzbehi\thanks{\href{mailto:rrouzbehi@chungbuk.ac.kr}{rrouzbehi@chungbuk.ac.kr}}\orcidlink{0009-0005-7139-5841}}

\affil[a]{Department of Physics, Semnan University, P.O. Box 35131-19111, Semnan, Iran}
\affil[b]{Department of Physics, Chungbuk National University, Cheongju, Chungbuk 28644, Korea}
\date{\today}

\begin{document}

\baselineskip 0.6 cm
\maketitle

\begin{abstract}
We study a classically scale-invariant extension of the Standard Model (SM) with four new fields in a hidden sector: a complex scalar $S$, two Dirac fermions $\psi^{1,2}$ and the gauge boson $V_\mu$ of a new $U(1)_D$ symmetry. The scalar $S$ couples to the SM through the Higgs portal. We identify regions of parameter space in which $V_\mu$, $\psi^1$ and $\psi^2$ are all stable, giving a three-component dark matter (DM) scenario. The model is classically scale invariant, and the electroweak symmetry breaks because of the loop effects. We impose the observed relic density and current direct detection bounds, and then study the finite temperature effective potential. For a benchmark point, we find a first-order phase transition and compute the resulting gravitational wave (GW) spectrum, which we compare with the projected sensitivities of space-based interferometers. Although the viable parameter space is small, it can be probed by future experiments.
\end{abstract}

\section{Introduction}\label{Introduction}
A wide range of astrophysical and cosmological observations provide compelling evidence for the existence of DM, which constitutes about 27 \% of the energy density of the Universe \cite{Bertone:2016nfn}. Despite its many successes, the SM contains no DM candidate. Therefore, a theory beyond the SM is necessary. Adding various degrees of freedom to the SM can introduce a candidate for DM. Weakly interacting massive particles (WIMPs) produced by thermal freeze-out are among the best-studied DM candidates~\cite{Feng:2022rxt}. Despite their popularity, no signal has been observed in direct detection experiments. Due to the strong constraints on direct detection experiments in one-component DM models, multi-component DM models seem more appropriate in some ways, in which Higgs-portal models being the most popular~\cite{PeymanZakeri:2018zaa, Gu:2013iy, Profumo:2009tb, Zurek:2008qg, Becker:2019xzq, Berman:2020kko, Yaguna:2021rds, Das:2022oyx, Costa:2022lpy, Khalil:2020syr, BasiBeneito:2022qxd, WileyDeal:2023trg, Bhattacharya:2016ysw, Bhattacharya:2018cgx, Bhattacharya:2022wtr, Ho:2022erb, DiazSaez:2021pmg, Yaguna:2021vhb, Hernandez-Sanchez:2020aop, Belanger:2020hyh, Yaguna:2019cvp, Elahi:2019jeo, Bernal:2018aon, YaserAyazi:2024dxk, Hosseini:2023qwu, Abkenar:2025urg, Abkenar:2024ket, Mohamadnejad:2021tke, YaserAyazi:2018lrv, Kim:2023onk, Neto:2026ijo, Abada:2026ons, Belyaev:2026umj, Sheikh:2026tjy, Ghorbani:2026inp, Duy:2026kxv, Escalona:2026trm, Qi:2026dwd, Chen:2025ihr, Qi:2025znq, Alvarado:2025vzv, Frank:2025jjt, Borah:2024emz}.

In the SM, the electroweak phase transition is a smooth crossover\cite{Kajantie:1996mn,Aoki:1999fi}. Extensions with extra scalars can make the electroweak phase transition strongly first order, as required for electroweak baryogenesis~\cite{Shaposhnikov:1987tw}; however, successful electroweak baryogenesis additionally requires appropriate CP violation and nonequilibrium transport. The present study does not calculate the baryon asymmetry. Moreover, such a transition also sources a stochastic GW background~\cite{Weir:2017wfa}. Therefore, the detection of these waves in various planned experiments could serve as evidence for new physics\cite{Crowder:2005nr, Kudoh:2005as, Yagi:2011wg}.
One of the fundamental challenges of particle physics is the hierarchy and fine-tuning problem. One approach to this problem is to forbid the Higgs mass term through classical scale invariance. In this paper, we present a classically scale-invariant extension of the SM. This framework provides a mechanism for generating the symmetry-breaking scales, but classical scale invariance alone does not establish a general solution to the hierarchy problem in an unspecified ultraviolet completion. Our
model is similar to that considered in\cite{Khoze:2022nyt, Kim:2019ogz}, although we have generalized the model to a three-component DM model by imposing numerous constraints; the parameter space of our model has become entirely different. The model includes four new fields: two fermions, a complex singlet scalar, and a vector field with $U_D(1)$ gauge symmetry. We probe the parameter space of the model according to constraints from relic density and direct detection. DM relic density is reported by the Planck collaboration\cite{Planck:2018vyg}, and the DM-nucleon cross section is constrained by XENONnT and LUX-ZEPLIN (LZ) experiment results\cite{XENON:2023cxc,LZ:2024zvo}. We investigate the possibility of the electroweak phase transition with respect to the bounded parameter space, where we use the effects of the effective potential at finite temperature. We identify points in the parameter space that satisfy the given constraints, and for a benchmark point within this space, we obtain a strong first-order phase transition and gravitational waves. We show that, although the model's parameter space is limited, it can still be investigated in future experiments.

The paper is organized as follows. Section.~\ref{Model} defines the model. In Section.~\ref{Relic Density}, we examine the model's parameter space in light of the observed relic density. In Section.~\ref{Direct Detection}, the results of the latest direct detection experiments are examined within the context of the model.
Electroweak phase transitions and gravitational wave signals are studied in Section.~\ref{sec:ewpt-gw}. Finally, our conclusion comes in Section.~\ref{conclusion}.

\section{The Model}\label{Model}
The model contains four new fields in which a vector field $V_{\mu}$ and two spinor fields $\psi^{1,2}$ can play the role of DM. A complex scalar,$S$, mediates between SM and the dark sector. The scalar and fermions are charged under a new dark $U(1)_D$ gauge symmetry, whose gauge boson is $V_\mu$. All of these fields are singlets under the SM gauge groups. We implement a discrete symmetry under which the new fields transform as follows
\begin{equation}\label{symmetry}
S\longrightarrow S^*,~~~  V_\mu \longrightarrow -V_\mu,~~~ \psi_L \longrightarrow -\psi_R,~~~~ \psi_R \longrightarrow -\psi_L,
\end{equation}
while all SM particles are even. The dark sector is invariant under the transformations of the $U_D(1)$ gauge group:
\begin{align} \label{2-2}
& \psi^a_L \rightarrow e^{i Q^a_L \alpha(x)} \psi^a_L, \nonumber \\
& \psi^a_R \rightarrow e^{i Q^a_R \alpha(x)} \psi^a_R, \nonumber \\
& S \rightarrow e^{i Q_S \alpha(x)} S, \nonumber \\
& V_{\mu} \rightarrow V_{\mu} - \frac{1}{g_v} \partial_{\mu}{\alpha(x)},
\end{align}
where the $U(1)_D$ charge of the new particles, $ Q_{L,R,S} $ are given in Table \ref{Table} and $a\in \{1,2\}$.
\begin{table}
\begin{center}
\begin{tabular}{| l | l | l | l | l | l | l |}
\hline
Field&$S$&$V_\mu$&$\psi^1_L$&$\psi^1_R$&$\psi^2_L$&$\psi^2_R$\\ \hline
$U(1)_D$ charge($Q$)&1&0&$\frac{1}{2}$&$-\frac{1}{2}$&$-\frac{1}{2}$&$\frac{1}{2}$\\ \hline
\end{tabular}
\end{center}
\caption{\label{Table}The charges of the dark sector particles under the new $U(1)_D$ symmetry.}
\end{table}
The most general Lagrangian by renormalizable interactions is given by:
\begin{align}
&{\cal L} ={\cal L}_{SM}-\frac{1}{4} V_{\mu \nu} V^{\mu \nu}+ (D_{\mu} S)^{*} (D^{\mu} S)- V(H,S) \nonumber \\+
& \sum_{a=1}^{2}(i\bar\psi^a_L  \gamma^{\mu}D_{\mu}\psi^a_L+ i\bar\psi^a_R  \gamma^{\mu}D_{\mu}\psi^a_R)
-g_{s,1} S\bar\psi^1_L \psi^1_R -g_{s,2} S^* \bar\psi^2_L \psi^2_R+H.C.,
\label{eq:lagrangian}
\end{align}
where $ {\cal L} _{SM} $ is the SM Lagrangian without the Higgs potential term. The covariant derivative is
\begin{align}
& D_{\mu}= (\partial_{\mu} + i Q g_{v} V_{\mu}), \quad \text{and}\nonumber \\
& V_{\mu \nu}= \partial_{\mu} V_{\nu} - \partial_{\nu} V_{\mu}.\end{align}
Imposing a dark charge-conjugation symmetry forbids kinetic mixing between $V_\mu$ and the hypercharge boson.
The most general scale-invariant potential $ V(H,S) $ that is renormalizable and invariant under gauge symmetry is
\begin{equation}
V(H,S) = \frac{1}{6} \lambda_{H} (H^{\dagger}H)^{2} + \frac{1}{6} \lambda_{S} (S^{*}S)^{2} + 2 \lambda_{S H} (S^{*}S) (H^{\dagger}H). \label{2-3}
\end{equation}
Note that the quartic portal interaction, $ \lambda_{S H} (S^{*}S) (H^{\dagger}H) $, is the only connection between the dark sector and the SM.
SM Higgs field $ H $ as well as the dark scalar $S$ can receive VEVs, respectively breaking the electroweak and $U_{D}(1)$ symmetries.
In unitary gauge, the imaginary component of $ S $ can be absorbed as the longitudinal part of $V_{\mu}$.
In this gauge, we can write
\begin{equation}
H = \frac{1}{\sqrt{2}} \begin{pmatrix}
0 \\ h_{1} \end{pmatrix} \, \, \, and \, \, \, S = \frac{1}{\sqrt{2}} h_{2} , \label{2-4}
\end{equation}
where $ h_{1} $ and $ h_{2} $ are real scalar fields which can receive VEVs. Now, the tree-level potential becomes
\begin{equation}
V^{tree} = \frac{1}{4 !} \lambda_{H} h_{1}^{4} + \frac{1}{4 !} \lambda_{S} h_{2}^{4} + \frac{1}{2} \lambda_{S H} h_{1}^{2} h_{2}^{2}. \label{2-5}
\end{equation}
Necessary and sufficient conditions for local minimum of $ V^{tree} $ in which  vacuum expectation values $ \langle h_{1} \rangle = \nu_{1} $ $\rm and$ $ \langle h_{2} \rangle = \nu_{2} $, can be written as:
\begin{align}
& \frac{\partial V^{tree}}{\partial h_{i}} \bigg\rvert_{\nu_{1},\nu_{2}} = 0 \label{2-7} \\
& \frac{\partial^{2} V^{tree}}{\partial h_{i} ^{2}} \bigg\rvert_{\nu_{1},\nu_{2}} > 0  \label{2-8} \\
& det(H (\nu_{1},\nu_{2})) > 0 , \label{2-9}
\end{align}
where $ det(H (\nu_{1},\nu_{2})) $ is determinant of the Hessian matrix. Condition (\ref{2-7}) for non-vanishing VEVs leads to the following constraints
\begin{align}
&\lambda_{H} \lambda_{S} = (3! \lambda_{S H})^{2}
&\frac{\nu_{1}^{2}}{\nu_{2}^{2}} = - \frac{3! \lambda_{S H}}{\lambda_{H}}. \label{2-10}
\end{align}
Conditions (\ref{2-7}) and (\ref{2-8}) require $ \lambda_{H} > 0 $, $ \lambda_{S} > 0 $, and $ \lambda_{S H} < 0 $. However, condition (\ref{2-9}) will not be satisfied, because $ det(H (\nu_{1},\nu_{2})) = 0 $. When the determinant of the Hessian matrix is zero, the second derivative test is inconclusive, and the point $ (\nu_{1},\nu_{2}) $ could be any of a minimum, maximum, or saddle point. However, in the model, constraint (\ref{2-10}) is defined as a flat direction, in which $ V^{tree} = 0 $. Thus, it is the stationary line or a local minimum. The important point is that for other directions $ V_{eff}^{1-loop(T=0)} > 0 $, and the tree-level potential only vanishes along the flat direction. Thus, the full potential of the theory will be dominated by higher-loop contributions along the flat direction and specifically by the one-loop effective potential. The one-loop correction, $V_{eff}^{1-loop}$, lifts the flat direction slightly and selects a minimum, which picks out a specific value along the ray as the minimum with $ V_{eff}^{1-loop} < 0 $ and vacuum expectation value $ \nu^{2} = \nu_{1}^{2} + \nu_{2}^{2} $ characterized by an RG scale $ \Lambda $. Since at the minimum of the one-loop effective potential $ V^{tree} \geqslant 0 $ and $ V_{eff}^{1-loop} < 0 $, the minimum of $ V_{eff}^{1-loop} $ along the flat direction (where $ V^{tree}=0 $) is a global minimum of the full potential, and so spontaneous symmetry breaking takes place. As a result, we suppose $ h_{1} \rightarrow \nu_{1} + h_{1} $ and $ h_{2} \rightarrow \nu_{2} + h_{2} $ and the electroweak symmetry breaks with  value $ \nu_{1} = 246 $ GeV. In the tree-level potential, since $ h_{1} $ and $ h_{2} $ mix with each other, they can be rewritten in terms of the mass eigenstates $ H_{1} $ and $ H_{2} $ as
\begin{equation}
\begin{pmatrix}
H_{1}\\H_{2}\end{pmatrix}
=\begin{pmatrix} cos \alpha~~~  -sin \alpha \\sin \alpha  ~~~~~cos \alpha
\end{pmatrix}\begin{pmatrix}
h_1 \\  h_{2}
\end{pmatrix}, \label{2-11}
\end{equation}
where $ H_{2} $ is along the flat direction, thus $ M_{H_{2}} = 0 $, and $ H_{1} $ is perpendicular to the flat direction, which we identify as the SM-like Higgs observed at the LHC with $ M_{H_{1}} = 125 $ GeV. After the symmetry breaking, we have the following constraints:
\begin{align}
& \nu_{2} =  \frac{M_{V}}{g_v} , &\nonumber
& sin \alpha =  \frac{\nu_{1}}{\sqrt{\nu_{1}^{2}+\nu_{2}^{2}}} \nonumber \\
&M_{\psi_1}= \frac{g_{s_1}M_{V}}{\sqrt{2}g_v}&\nonumber
&M_{\psi_2}= \frac{g_{s_2}M_{V}}{\sqrt{2}g_v}&\nonumber
& \lambda_{H} =  \frac{3 M_{H_{1}}^{2}}{ \nu_{1}^{2}} cos^{2} \alpha  \nonumber  \\
& \lambda_{S} =  \frac{3 M_{H_{1}}^{2}}{ \nu_{2}^{2}} sin^{2} \alpha  &\nonumber
& \lambda_{S H} =  - \frac{ M_{H_{1}}^{2}}{2 \nu_{1} \nu_{2} } sin \alpha \, cos \alpha , \\ \label{2-12}
\end{align}
where $M_{\psi_{1,2}}$ and $M_{V}$ are the masses of vector and fermion fields after symmetry breaking.
Conditions (\ref{2-12}) constrain the free parameters of the model to four independent parameters.
We choose $M_{V}$,  $M_{\psi_{1}} $, $M_{\psi_{2}}$ and  $g_v$ as the independent parameters of the model.

Since, at tree-level, $M_{H_{2}}=0$, the elastic scattering cross section of DM off nuclei becomes severely large, and the model would be excluded by direct detection experiments. However, the radiative corrections give a mass to the massless eigenstate $ H_{2} $. One-loop corrections to the potential, via the Gildener–Weinberg formalism \cite{PhysRevD.13.3333}, shift the scalon mass to values that can be even higher than the SM Higgs mass. Along the flat direction, the one-loop effective potential has the general form \cite{PhysRevD.13.3333}
\begin{equation}
V_{eff}^{1-loop(T=0)} = a H_{2}^{4} + b H_{2}^{4} \, \ln \frac{H_{2}^{2}}{\Lambda^{2}}  , \label{2-13}
\end{equation}
where  $ a $ and $ b $ are the dimensionless constants that given by
\begin{align}
& a =  \frac{1}{64 \pi^{2} \nu^{4}}  \sum_{k=1}^{n} g_{k}  M_{k}^{4} \ln \frac{M_{k}^{2}}{\nu^{2}}  , \nonumber \\
& b = \frac{1}{64 \pi^{2} \nu^{4}} \sum_{k=1}^{n} g_{k}  M_{k}^{4} . \label{2-14}
\end{align}
In (\ref{2-14}), $ M_{k} $ and $ g_{k} $ are the  tree-level mass and the internal degrees of freedom of the particle $ k $, respectively (In our convention $ g_{k} $
takes positive values for bosons and negative values for fermions).
Minimizing (\ref{2-13}) shows that the potential has a non-trivial stationary point at a value of the RG scale $ \Lambda $, given by
\begin{equation}
\Lambda = \nu \exp \left( \frac{a}{2b} + \frac{1}{4} \right)   , \label{2-15}
\end{equation}
Eq.~(\ref{2-15}) can now be used to find the form of the one-loop effective potential along the flat direction in terms of the one-loop VEV $ \nu $
\begin{equation}
V_{eff}^{1-loop(T=0)} = b H_{2}^{4} \, \left( \ln \frac{H_{2}^{2}}{\nu^{2}} - \frac{1}{2} \right) , \label{2-16}
\end{equation}
It is remarkable that the scalon does not remain massless beyond the tree approximation. Regarding $ V_{eff}^{1-loop(T=0)} $, $ M_{H_{2}} $  will be
\begin{equation}
M_{H_{2}}^{2} = \frac{d^2 V_{eff}^{1-loop(T=0)}}{d H_{2}^{2}} \bigg\rvert_{\nu} = 8 b \nu^{2} . \label{2-17}
\end{equation}
Considering (\ref{2-14}),  the scalon mass can be expressed in terms of other particle masses
\begin{equation}
M_{H_{2}}^{2} = \frac{1}{8 \pi^{2} \nu^{2}} \left( M_{H_{1}}^{4} + 6  M_{W}^{4} + 3  M_{Z}^{4} + 3  M_{V}^{4} - 12 M_{t}^{4} -4M_{\psi_{1}}^{4} -4M_{\psi_{2}}^{4} \right) . \label{2-18}
\end{equation}
where $ M_{W,Z,t} $ are the masses of the W, Z gauge bosons, and top quark, respectively. As mentioned before, $ M_{H_{1}} = 125 $ GeV and $\nu^{2} = \nu_{1}^{2} + \nu_{2}^{2} $.
 From (\ref{2-18}), we have a constraint on the parameter space of our model where  $ M_{H_{2}}>0$.

Note that according to (\ref{2-18}) and (\ref{2-12}), $ M_{H_{2}} $ is completely determined by the independent parameters of the model, i.e., $ M_{V} $, $M_{\psi_{1}} $, $M_{\psi_{2}}$ and the coupling $ g_v $. If $M_V< 2M_{\psi_{1}} $ and $M_V< 2M_{\psi_{2}} $, all three $V_{\mu}$, $\psi_1$ and $\psi_2$ fields are stable DM candidates.

 \section{Relic Density}\label{Relic Density}
\begin{figure}
	\begin{center}
		\centerline{\hspace{0cm}\epsfig{figure=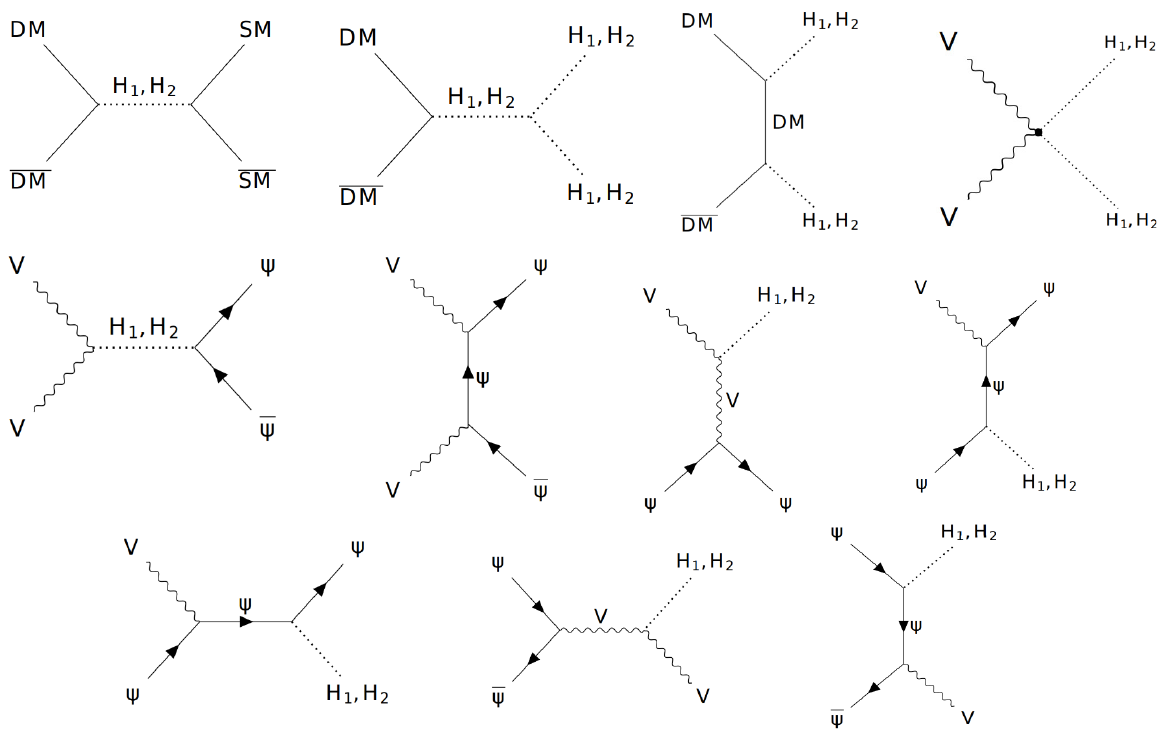,width=15cm}}
		\centerline{\vspace{-0.3cm}}
		\caption{ Relevant Feynman diagrams for relic density. SM and DM stand for Standard Model massive fermions and gauge bosons and dark matter particles.
The field $\psi$ can correspond to either $\psi^1$
 or $\psi^2$.} \label{feynmanfig}
	\end{center}
\end{figure}

 Assuming only 2 → 2 reactions, the Boltzmann equation describing the evolution of the abundance of the $i$-th DM candidate reads\cite{Alguero:2023zol}
 \begin{equation}
 \frac{dn_i}{dt}=- \sum_{\alpha \leq \beta ; \gamma \leq \delta} C_{\alpha\beta}n_\alpha n_\beta \langle\upsilon \sigma_{\alpha\beta \rightarrow \gamma\delta}\rangle (\delta_{i\alpha}+\delta_{i\beta}-\delta_{i\gamma}-\delta_{i\delta})-3H(T)n_i ,
 \end{equation}
 where $n_i$ with $i = 1, 2, 3$ is the number density for each DM species, $C_{\alpha\beta}$ is a combinatoric factor, $C_{\alpha\beta}$ = 1/2 if $\alpha=\beta$ and 1 otherwise, and $H(T)$ is the Hubble expansion rate. Greek indices run over 0, 1, 2, 3, with zero referring to the sector that includes all
SM particles and $H_2$ particle. Considering the conservation of entropy, $\frac{ds}{dt}=-3Hs$, we get
 \begin{equation}
 3H\frac{dY_i}{ds}=\sum_{\alpha \leq \beta ; \gamma \leq \delta}C_{\alpha\beta}Y_\alpha Y_\beta \langle\upsilon \sigma_{\alpha\beta \rightarrow \gamma\delta}\rangle (\delta_{i\alpha}+\delta_{i\beta}-\delta_{i\gamma}-\delta_{i\delta}),
 \end{equation}
 where $Y_i = n_i/s$.  By solving these Boltzmann equations, we can determine the relic densities, $\Omega_{V, \psi_1 ,\psi_2}$ .
 In Figure \ref{feynmanfig}, all relevant Feynman diagrams for calculating the DM relic density are plotted, including all annihilation, semi-annihilation, and conversion processes. The range of independent model parameters for scanning the results is as follows:
 \begin{itemize}
 	\item The DM mass $M_V$ is between $1-5000~\rm GeV$;
    \item The DM mass $M_{\psi_{1,2}}$ is between $1-5000~\rm GeV$;
 	\item The $ g_v $ is between $1-3$;
 \end{itemize}

We also used constraints related to mixing between scalars ($sin \alpha$) in checking our results. Non-observation of DM at the LHC corresponds to an upper bound on the value of $sin \alpha$. Higgs signal strength measurement implies a bound on the Higgs mixing angle as $sin \alpha < 0.23$ \cite{ATLAS:2016neq,CMS:2022dwd,Khan:2025yko}. We used the micrOMEGAs package\cite{Alguero:2023zol} to numerically solve coupled Boltzmann differential equations. According to the data from the Planck Collaboration\cite{Planck:2018vyg}, the DM
constraint in this model reads
\begin{equation}
\Omega_{DM} h^{2} = \Omega_{V} h^{2}+ \Omega_{{\psi_1}} h^{2}+ \Omega_{{\psi_2}} h^{2} = 0.120 \pm 0.001.
\end{equation}
We also define the fraction of the DM density of each component by
\begin{equation}
\xi_V= \frac{\Omega_V}{\Omega_{DM}}, ~~~~~ \xi_{\psi_1}= \frac{\Omega_{{\psi_1}}}{\Omega_{DM}}, ~~~~~ \xi_{\psi_2}= \frac{\Omega_{{\psi_2}}}{\Omega_{DM}},~~~ \xi_V+\xi_{\psi_1}+\xi_{\psi_2} =1.
\end{equation}

\begin{figure}
	\begin{center}
		\centerline{\hspace{0.1cm}\epsfig{figure=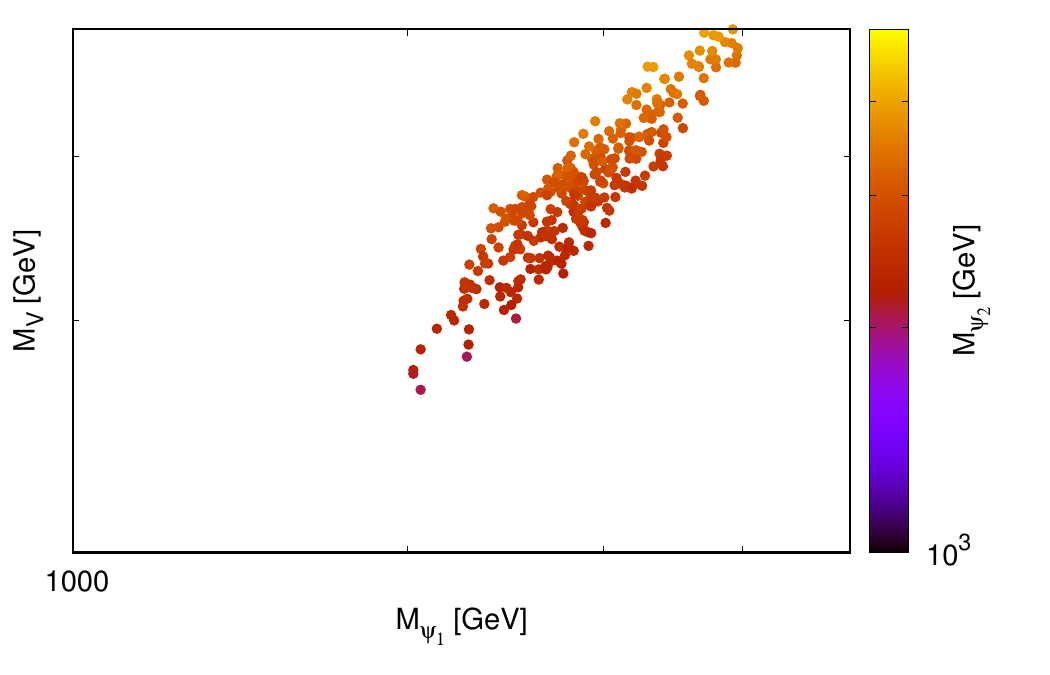,width=13cm}}
        \centerline{\vspace{0.1cm}}
		\caption{ The allowed range of parameter space consistent with the DM relic density.} \label{relic}
	\end{center}
\end{figure}

 Figure \ref{relic} shows the allowed parameter space that is consistent with the relic density value reported by the Planck Collaboration\cite{Planck:2018vyg}.

  \section{Direct Detection}\label{Direct Detection}
\begin{figure}
	\begin{center}
		\centerline{\hspace{-4cm}\epsfig{figure=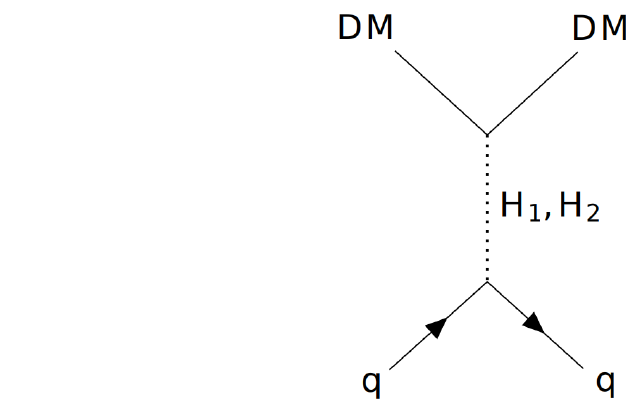,width=8cm}}
        \centerline{\vspace{-0.2cm}}
		\caption{ Relevant Feynman diagram for direct detection.} \label{direct}
	\end{center}
\end{figure}

\begin{figure}
	\begin{center}
		\centerline{\hspace{0cm}\epsfig{figure=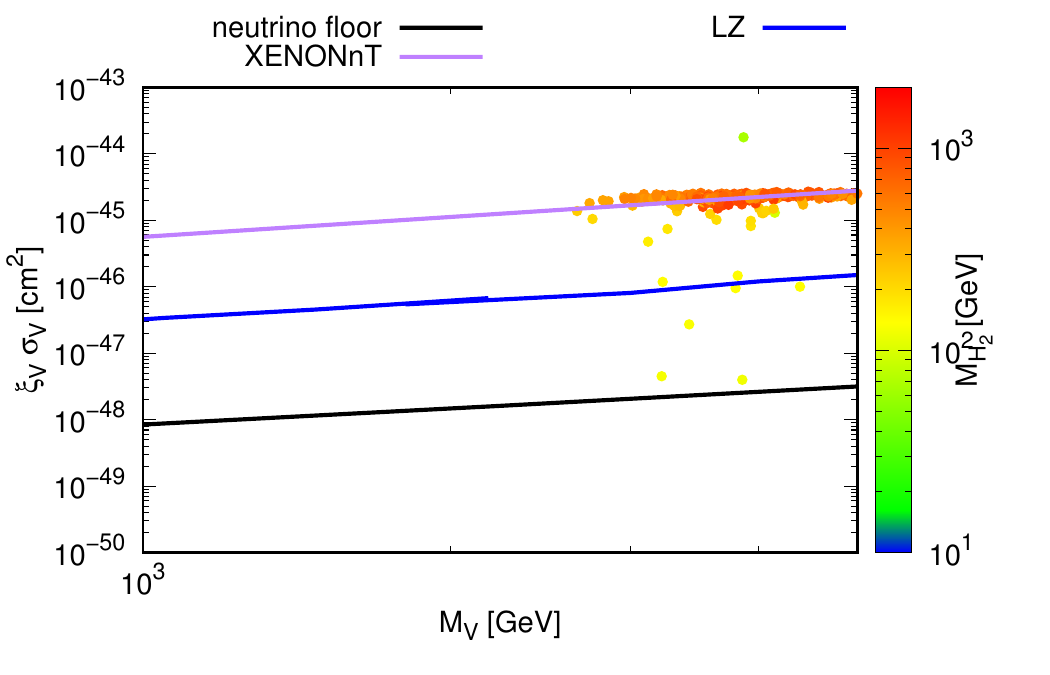,width=9cm}\hspace{0.5cm}\hspace{0cm}\epsfig{figure=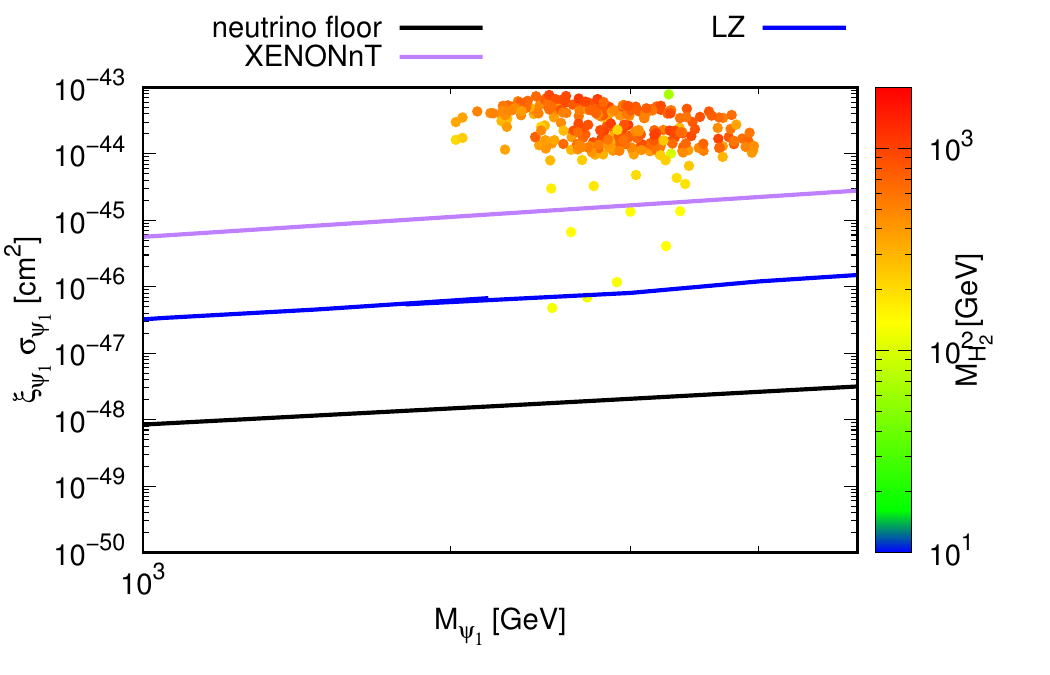,width=9cm}}
		\centerline{\vspace{0.2cm}\hspace{0.3cm}(a)\hspace{8cm}(b)} \centerline{\hspace{0cm}\epsfig{figure=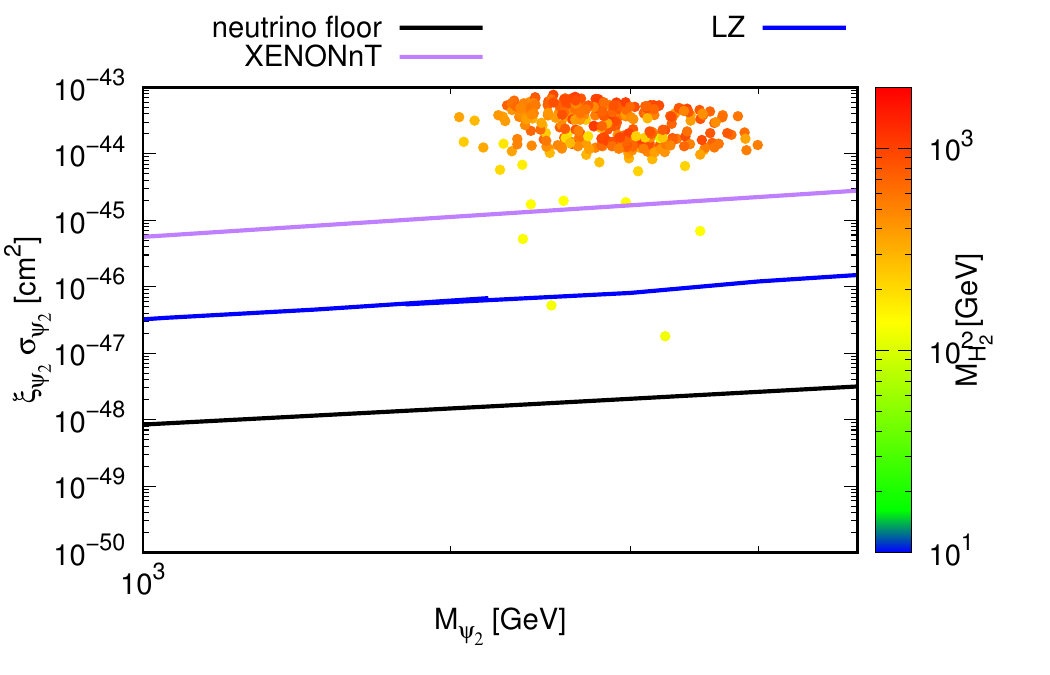,width=9cm}\hspace{0.5cm}\hspace{0cm}}		
		\centerline{\vspace{0.2cm}\hspace{0.6cm}(c)}
		\centerline{\vspace{-0.2cm}}
		\caption{The allowed range of parameter space consistent with DM relic density and DD.  In (a) $\xi_V \sigma_V$ VS $M_V$ and in (b) $\xi_{\psi_1} \sigma_{\psi_1}$ VS $M_{\psi_1}$ and in (c) $\xi_{\psi_2} \sigma_{\psi_2}$ VS $M_{\psi_2}$ has shown.} \label{direct detection}
	\end{center}
\end{figure}

 In our model, the DM particles interact with quarks via the exchange of $H_1$ or
$H_2$ within the Higgs portal framework, leading to a spin-independent cross section for DM-nucleon interactions(see figure \ref{direct}).
 The spin-independent DM-nucleon cross sections for spinor and vector DM are given by

 \begin{align}
& \sigma_{\psi_{1}-N} ^{SI} =\frac{g_{s_1}^2 \sin ^2\alpha \cos ^2\alpha}{2 \nu_1^2}(\frac{1}{M_{H1} ^2}-\frac{1}{M_{H2} ^2})^2 \frac{M_N ^4 M_{\psi_1}^2}{\pi (M_N +M_{\psi_1})^2}f_N ^2  , \nonumber \\
& \sigma_{\psi_{2}-N} ^{SI} =\frac{g_{s_2}^2 \sin ^2\alpha \cos ^2\alpha}{2 \nu_1^2}(\frac{1}{M_{H1} ^2}-\frac{1}{M_{H2} ^2})^2 \frac{M_N ^4 M_{\psi_2}^2}{\pi (M_N +M_{\psi_2})^2}f_N ^2 ,\nonumber \\
&  \sigma_{V-N} ^{SI} = g_v^4 \cos ^4\alpha (\frac{1}{M_{H1} ^2}-\frac{1}{M_{H2} ^2})^2 \frac{M_N ^4 }{\pi (M_N +M_V)^2}f_N ^2 , \label{cross section}
\end{align}
 where $M_N$ represents the mass of the nucleon and $f_N$ = 0.3 characterizes the coupling
between the Higgs and nucleons. Assuming that the share of each DM candidate in the local DM
density mirrors their contribution to the relic density, many researchers have imposed
constraints on the rescaled DM-nucleon cross sections, namely $\xi_V \sigma_V$, $\xi_{\psi_1}\sigma_{\psi_1}$, and $\xi_{\psi_2}\sigma_{\psi_2}$,
based on experimental findings. In Fig.~\ref{direct detection}, rescaled DM-Nucleon cross sections are depicted for the parameters that are in agreement with the relic density. We use the results of the XENONnT and LZ experiments\cite{XENON:2023cxc,LZ:2024zvo}. The LZ experiment places the strongest possible constraint on direct detection results. The neutrino floor, which is an irreducible background resulting from the scattering of SM neutrinos by nuclei, is also considered\cite{Billard:2013qya}.

An important point is that the phenomenology of the model is subject to several fundamental constraints. In the modeling section, we encounter the following constraints:

\begin{equation}
M_V< 2M_{\psi_{1}}, ~~~~~ M_V< 2M_{\psi_{2}}, ~~~~~ M_{H_{2}}>0.
\end{equation}

In the experimental sector, we also encounter the mixing angle constraint($sin \alpha$) and a very stringent LZ constraint. For this reason, our parameter space is limited; nevertheless, there still exist points within this parameter space that satisfy all the aforementioned constraints.

\section{Electroweak phase transition and gravitational wave signal}\label{sec:ewpt-gw}
In the previous sections, we studied the relic abundance and direct detection properties of the three DM candidates. We now investigate the thermal evolution of the model and the gravitational waves (GWs) that may be produced during symmetry breaking. The Higgs portal connects the electroweak and dark scalar sectors, while the vector $V_\mu$ and the two fermions $\psi_1$ and $\psi_2$ obtain their masses from the dark scalar vacuum expectation value. Therefore, the parameters that determine the DM phenomenology also enter the phase transition analysis. We follow the finite-temperature approach of Ref.~\cite{YaserAyazi:2024dxk} and apply it to the particle content of the present model.
\subsection{Effective potential at finite temperature}\label{subsec:thermal-potential}
To study the GW resulting from the electroweak phase transition, we need the effective potential as a function of the temperature and the scalar background. In addition to the 1-loop zero-temperature potential (\ref{2-16}), we can also consider the 1-loop corrections at finite temperature in the effective potential. The one-loop thermal contribution is given by~\cite{PhysRevD.9.3320}
\begin{equation}
 V_{\rm eff}^{1-loop(T\neq 0)}(H_2,T)=\frac{T^4}{2\pi^2}
 \sum_k g_k J_{B,F}\!\left(\frac{M_kH_2}{\nu T}\right),
 \label{eq:thermal-one-loop}
\end{equation}
where the bosonic and fermionic functions are
\begin{equation}
 J_{B,F}(x)=\int_0^\infty dy\,y^2
 \ln\!\left(1\mp e^{-\sqrt{y^2+x^2}}\right).
 \label{eq:thermal-functions}
\end{equation}
The upper sign corresponds to bosons and the lower sign to fermions. Here, the argument is $x=m_k(H_2)/T$, and the degrees of freedom are signed: $g_{H_1}=1$, $g_W=6$, $g_Z=g_V=3$, $g_t=-12$, and $g_{\psi_1}=g_{\psi_2}=-4$. The minus sign for fermions is therefore already included in $g_k$. The tree-level mass of the scalon is zero; its
field-independent contribution to $ V_{\rm eff}^{1-loop(T\neq 0)}$ disappears when the potential is normalized at $H_2=0$. The thermal functions can also be evaluated through modified Bessel functions. The short-sum approximations used in Refs.~\cite{YaserAyazi:2024dxk, Mohamadnejad:2019vzg, PhysRevD.13.3333, PhysRevD.9.3320} are
\begin{equation}
 \begin{aligned}
 J_B(x)&\simeq-\sum_{\ell=1}^{3}\frac{x^2}{\ell^2}K_2(\ell x),\\
 J_F(x)&\simeq-\sum_{\ell=1}^{2}
              \frac{(-1)^\ell x^2}{\ell^2}K_2(\ell x),
 \end{aligned}
 \label{eq:thermal-bessel}
\end{equation}
where $K_2$ is the modified Bessel function of the second kind. These finite sums approximate the integral definitions in Eq.~\eqref{eq:thermal-functions}; the complete Bessel representations are obtained by extending the sums to infinity. We also include the daisy contribution, which resums thermal corrections to the bosonic zero modes~\cite{Carrington:1991hz, Arnold:1992rz}:
\begin{equation}
 V_{\rm daisy}(H_2,T)=\frac{T}{12\pi}
 \sum_b n_b\left[m_b^3(H_2)-\overline m_b^{\,3}(H_2,T)\right].
 \label{eq:daisy}
\end{equation}
The sum contains the scalar modes and longitudinal gauge modes, with positive multiplicities $n_b$. The quantities $m_b^2$ and $\overline m_b^{\,2}$ are the eigenvalues before and after adding the thermal self-energy to the corresponding mass matrix. In the thermal prescription adopted here, the self-energies are
\begin{equation}
 \begin{gathered}
 \Pi_W=\frac{11}{6}g_{\rm SM}^2T^2,
 \qquad \Pi_V=\frac{2}{3}g_v^2T^2,\\
 \Pi_{(W^3,B)}=\frac{11T^2}{6}
   \begin{pmatrix}g_{\rm SM}^2&0\\0&g_{\rm SM}^{\prime\,2}\end{pmatrix},
 \qquad
 \Pi_{(h_1,h_2)}=T^2
   \begin{pmatrix}c_1&0\\0&c_2\end{pmatrix},\\
 \begin{aligned}
 c_1&=\frac{\lambda_H}{24}+\frac{\lambda_{SH}}{12}
       +\frac{3g_{\rm SM}^2+g_{\rm SM}^{\prime\,2}}{16}
       +\frac{\lambda_t^2}{4},\\
 c_2&=\frac{\lambda_S}{24}+\frac{\lambda_{SH}}{12}
       +\frac{g_v^2}{4}+\frac{g_{s1}^2+g_{s2}^2}{12}.
 \end{aligned}
 \end{gathered}
 \label{eq:thermal-self-energies}
\end{equation}
Here, $g_{\rm SM}$ and $g'_{\rm SM}$ denote the electroweak gauge couplings, and $\lambda_t$ is the top Yukawa coupling. The scalar and neutral gauge
matrices in Eq.~\eqref{eq:thermal-self-energies} are written in the $(h_1,h_2)$ and $(W^3,B)$ bases. We add these matrices to the tree-level mass matrices before diagonalization. The longitudinal gauge contribution includes two charged $W$ modes, two neutral modes, and one dark-vector mode. Although the fermions do not appear directly in the daisy sum, both dark Yukawa couplings contribute to $c_2$. Thus, the two fermions modify the thermal potential through Eq.~\eqref{eq:thermal-one-loop} and
through the scalar thermal mass. Combining the zero-temperature, thermal, and daisy terms, we obtain
\begin{equation}
 V_{\rm eff}(H_2,T)=V_{\rm eff}^{\rm 1-loop(T=0)}(H_2)+ V_{\rm eff}^{1-loop(T\neq 0)}(H_2,T)+V_{\rm daisy}(H_2,T).
 \label{eq:full-effective-potential}
\end{equation}
For convenience, we measure the potential relative to the symmetric configuration at the same temperature:
\begin{equation}
  V_{\rm eff}(H_2,T) \to V_{\rm eff}(H_2,T)-V_{\rm eff}(0,T).
 \label{eq:normalized-potential}
\end{equation}
This choice gives $V_{\rm eff}(0,T)=0$ without changing the positions of the extrema or the free-energy difference between the phases.
All three contributions in Eq.~\eqref{eq:full-effective-potential} are retained in the phase transition analysis. Their dependence on $M_V$, $M_{\psi_1}$, $M_{\psi_2}$, and $g_v$ determines the thermal history of the three-component model.
\subsection{First-order phase transition and gravitational wave spectrum}\label{subsec:transition-and-gw}
A first-order transition can occur when the effective potential develops two local minima separated by a barrier. The symmetric configuration is located at $H_2=0$, whereas the broken phase has a nonzero scalar background. The critical temperature $T_C$ is the temperature at which the two phases have equal free energy. Denoting the broken minimum at this temperature by $\nu_C$, we determine $T_C$ and $\nu_C$ from
\begin{equation}
 V_{\rm eff}(\nu_C,T_C)=0,
 \qquad
 \left.\frac{\partial V_{\rm eff}(H_2,T_C)}{\partial H_2}
 \right|_{H_2=\nu_C}=0,
 \label{eq:critical-temperature}
\end{equation}
with both stationary points required to be local minima. The ratio $\nu_C/T_C$ describes the displacement along the flat direction. The usual approximate criterion for a strong electroweak transition is~\cite{Shaposhnikov:1987tw}
\begin{equation}
 \frac{\nu_C}{T_C}>1.
 \label{eq:ew-strength}
\end{equation}
\begin{figure}[h]
	\begin{center}
		\centerline{\hspace{0.1cm}\epsfig{figure=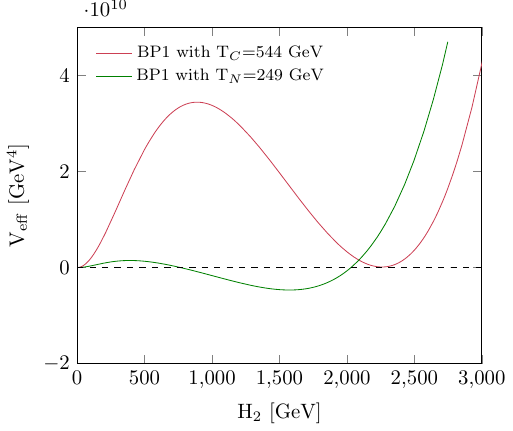,width=11cm}}
        \centerline{\vspace{0.1cm}}
		\caption{ Potential behavior for critical temperature and nucleation temperature.} \label{critical}
	\end{center}
\end{figure}

\begin{figure}[t!]
\centering
\includegraphics[width=0.7\linewidth]{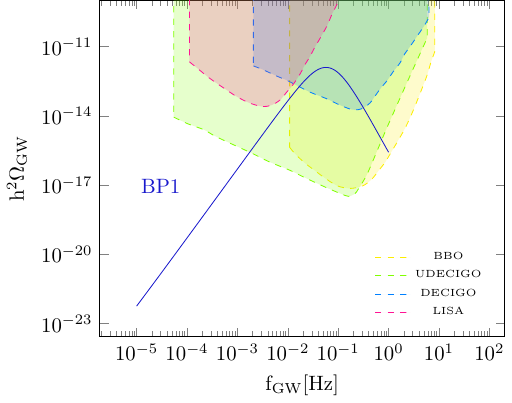}
\caption{GW spectrum for benchmark points of the Table \ref{table2}.} \label{GWs}
\end{figure}

\begin{figure}[h!]
 \centering
\includegraphics[]{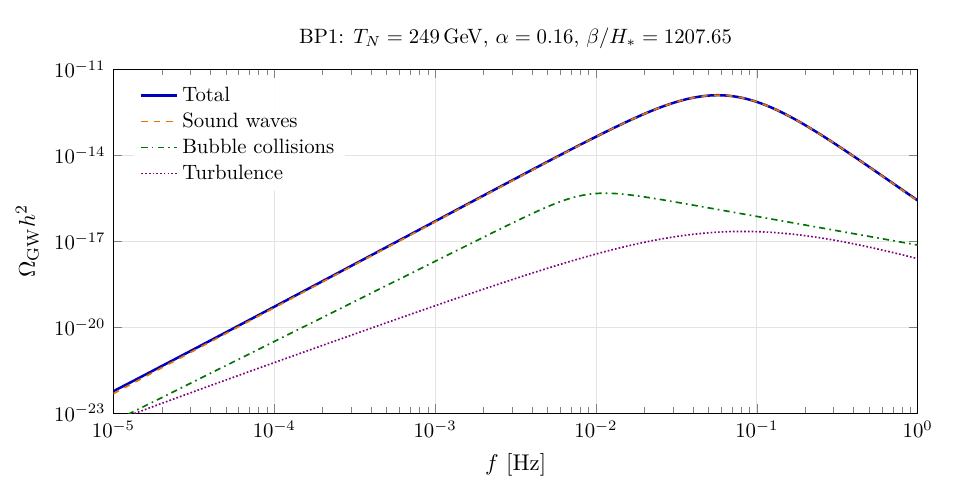}
 \caption{Present-day GW spectrum for the BP1 parameters in Table~\ref{table2}. The total includes the collision, acoustic,
 and turbulent fits specified in the text. The acoustic contribution controls the maximum. The curves use $g_*=100$, $v_w=1$, and the
 long-lived sound-wave approximation.}
 \label{fig:gw-bp1}
\end{figure}

To provide a clear understanding of the phase transition behavior, Fig.~\ref{critical} illustrates the effective potential for the chosen benchmark points at both the critical and nucleation temperatures. This criterion refers to the Higgs-doublet component. A precise assessment of baryon-number preservation requires the broken-phase sphaleron rate, which is beyond the GW estimate considered here. Below $T_C$, the broken minimum becomes energetically favored, but the barrier can delay the transition. Thermal fluctuations then produce bubbles
of the broken phase inside the symmetric plasma. The tunneling rate is exponentially controlled by $S_3(T)/T$, where the three-dimensional action for an $O(3)$-symmetric bubble is
\begin{equation}
 S_3(T)=4\pi\int_0^\infty dr\,r^2
 \left[\frac12\left(\frac{dH_2}{dr}\right)^2
       +V_{\rm eff}(H_2,T)\right].
 \label{eq:bounce-action}
\end{equation}
The radial field profile is obtained by solving
\begin{equation}
 \frac{d^2H_2}{dr^2}+\frac{2}{r}\frac{dH_2}{dr}
 =\frac{\partial V_{\rm eff}(H_2,T)}{\partial H_2},
 \label{eq:bounce-equation}
\end{equation}
with
\begin{equation}
 \left.\frac{dH_2}{dr}\right|_{r=0}=0,
 \qquad H_2(r\rightarrow\infty)=0.
 \label{eq:bounce-boundaries}
\end{equation}
We use the \texttt{AnyBubble} package~\cite{Masoumi:2017trx} for the numerical bounce calculation. The nucleation temperature $T_N$ is estimated through $S_3(T_N)/T_N\simeq140$, as commonly used for electroweak-scale transitions in a radiation-dominated
Universe~\cite{Apreda:2001us}. It marks the onset of appreciable bubble nucleation; the subsequent completion of the transition also depends on bubble growth and percolation.
The released energy is transferred to the bubble walls and the surrounding plasma. Collisions of the scalar-field configurations, the acoustic motion of the fluid, and magnetohydrodynamic turbulence can each generate a stochastic GW background. We describe the present spectrum as the sum of these contributions~\cite{Caprini:2015zlo}:
\begin{equation}
 \Omega_{\rm GW}(f)h^2\simeq
 \Omega_{\rm coll}(f)h^2+\Omega_{\rm sw}(f)h^2
 +\Omega_{\rm turb}(f)h^2.
 \label{eq:gw-total}
\end{equation}
We approximate the GW production temperature by $T_N$ and assume radiation-dominated, adiabatic expansion after the transition.
In addition to $T_N$, the spectrum depends on the transition strength $\alpha$, the inverse duration $\beta/H_*$, and the wall velocity $v_w$.
We use the same energy-density definition of $\alpha$ as in Ref.~\cite{YaserAyazi:2024dxk}:
\begin{equation}
 \alpha=\frac{1}{\rho_*}
 \left.\Delta\!\left(V_{\rm eff}-T\frac{\partial V_{\rm eff}}{\partial T}
 \right)\right|_{T_N},
 \label{eq:transition-alpha}
\end{equation}
where $\Delta X=X_{\rm false}-X_{\rm true}$ and
\begin{equation}
 \rho_* = \frac{\pi^2g_*}{30}T_N^4.
 \label{eq:radiation-density}
\end{equation}
The parameter $g_*$ is the effective number of relativistic energy degrees of freedom. The second transition parameter is
\begin{equation}
 \frac{\beta}{H_*}
 =T_N\left.\frac{d}{dT}\left(\frac{S_3(T)}{T}\right)\right|_{T_N},
 \label{eq:beta-over-H}
\end{equation}
where $H_*=H(T_N)$. A larger value corresponds to a shorter characteristic transition time. The wall velocity is affected by the pressure difference
and the interaction of the wall with the plasma~\cite{Bodeker:2009qy}. For the benchmark spectrum, we take the relativistic-wall approximation
$v_w=1$ and set $g_*=100$.
For the bubble-collision contribution, we adopt the envelope-approximation fit~\cite{Huber:2008hg}
\begin{equation}
 \Omega_{\rm coll}(f)h^2=1.67\times10^{-5}
 \left(\frac{\beta}{H_*}\right)^{-2}
 \left(\frac{\kappa\alpha}{1+\alpha}\right)^2
 \left(\frac{g_*}{100}\right)^{-1/3}
 \frac{0.11v_w^3}{0.42+v_w^2}\,S_{\rm coll}(f),
 \label{eq:gw-collisions}
\end{equation}
where the spectral function and its peak frequency are
\begin{equation}
 S_{\rm coll}(f)=\frac{3.8(f/f_{\rm coll})^{2.8}}
                         {1+2.8(f/f_{\rm coll})^{3.8}},
 \label{eq:shape-collisions}
\end{equation}
\begin{equation}
 f_{\rm coll}=1.65\times10^{-5}
 \frac{0.62}{v_w^2-0.1v_w+1.8}\frac{\beta}{H_*}
 \left(\frac{T_N}{100\,\mathrm{GeV}}\right)
 \left(\frac{g_*}{100}\right)^{1/6}\mathrm{Hz}.
 \label{eq:frequency-collisions}
\end{equation}
The expansion and overlap of the bubbles also induce bulk fluid motion. The acoustic part of this motion sources GWs after the walls have
collided. Its contribution is described by~\cite{Hindmarsh:2015qta}
\begin{equation}
 \Omega_{\rm sw}(f)h^2=2.65\times10^{-6}
 \left(\frac{\beta}{H_*}\right)^{-1}
 \left(\frac{\kappa_v\alpha}{1+\alpha}\right)^2
 \left(\frac{g_*}{100}\right)^{-1/3}v_w S_{\rm sw}(f),
 \label{eq:gw-sound}
\end{equation}
with
\begin{equation}
 S_{\rm sw}(f)=\left(\frac{f}{f_{\rm sw}}\right)^3
 \left[\frac{7}{4+3(f/f_{\rm sw})^2}\right]^{7/2},
 \label{eq:shape-sound}
\end{equation}
\begin{equation}
 f_{\rm sw}=1.9\times10^{-5}\frac{1}{v_w}\frac{\beta}{H_*}
 \left(\frac{T_N}{100\,\mathrm{GeV}}\right)
 \left(\frac{g_*}{100}\right)^{1/6}\mathrm{Hz}.
 \label{eq:frequency-sound}
\end{equation}
We use the long-lived sound-wave approximation in Eq.~\eqref{eq:gw-sound}. A shorter acoustic lifetime would reduce this contribution~\cite{Ellis:2020awk}.
The turbulent component is estimated using~\cite{Caprini:2009yp}
\begin{equation}
 \Omega_{\rm turb}(f)h^2=3.35\times10^{-4}
 \left(\frac{\beta}{H_*}\right)^{-1}
 \left(\frac{\kappa_{\rm turb}\alpha}{1+\alpha}\right)^{3/2}
 \left(\frac{g_*}{100}\right)^{-1/3}v_w S_{\rm turb}(f),
 \label{eq:gw-turbulence}
\end{equation}
where
\begin{equation}
 S_{\rm turb}(f)=\frac{(f/f_{\rm turb})^3}
 {(1+f/f_{\rm turb})^{11/3}(1+8\pi f/h_*)}.
 \label{eq:shape-turbulence}
\end{equation}
We retain the characteristic frequency by
\begin{equation}
 f_{\rm turb}=2.27\times10^{-5}\frac{1}{v_w}\frac{\beta}{H_*}
 \left(\frac{T_N}{100\,\mathrm{GeV}}\right)
 \left(\frac{g_*}{100}\right)^{1/6}\mathrm{Hz},
 \label{eq:frequency-turbulence}
\end{equation}
whereas the inverse Hubble time redshifted to today is
\begin{equation}
 h_*=1.65\times10^{-5}
 \left(\frac{T_N}{100\,\mathrm{GeV}}\right)
 \left(\frac{g_*}{100}\right)^{1/6}\mathrm{Hz}.
 \label{eq:redshifted-Hubble}
\end{equation}
\begin{table}[h!]
\centering 
\begin{tabular}{l c c rrrrrrr}
\hline\hline
 $\#$ &$M_V (GeV)$ &$M_{\psi_1}(GeV)$ &$M_{\psi_2}(GeV)$&$g_v$&$g_{s_1}$&$g_{s_2}$&$M_{H_2}(GeV)$\\
\hline
1&3216&2514&2510&2.4&2.65&2.64&127.5 \\
\hline\hline 
$\#$ &$\Omega_{V} h^{2}$ &$\Omega_{{\psi_1}} h^{2}$ &$\Omega_{{\psi_2}} h^{2} $&$\Omega_{DM} h^2$&$\xi_V \sigma_V (cm^2)$&$\xi_{{\psi_1}}\sigma_{{\psi_1}} (cm^2)$&$\xi_{{\psi_2}}\sigma_{{\psi_2}} (cm^2)$\\
\hline
1&$2.98\times10^{-3}$&$5.199\times10^{-2}$&$5.7\times10^{-2}$&$1.2\times10^{-1}$&$4.5\times10^{-48}$&$4.79\times10^{-47}$&$5.24\times10^{-47}$\\
\hline\hline
$\#$ &$T_C (GeV)$&$T_N (GeV)$&$\alpha$&$\beta/H_\ast$&$(\Omega_{GW} h^2)_{max}$\\
\hline
1&544&249&0.16&1207.65&$1.25\times 10^{-12}$\\
\hline
\end{tabular}
\caption{\label{table2}The benchmark point with DM and phase transition parameters.} 
\end{table}
The scale $f_{\rm turb}$ need not coincide with the maximum of the turbulence spectrum because of its additional $h_*$ dependence. The coefficient in Eq.~\eqref{eq:frequency-turbulence} follows the earlier paper; the commonly used normalization in Ref.~\cite{Caprini:2009yp} is
$2.7\times10^{-5}\,\mathrm{Hz}$. The factors $\kappa$, $\kappa_v$, and $\kappa_{\rm turb}$ specify the fractions of released energy assigned to scalar-field gradients, bulk
fluid motion, and turbulence. To compare directly with the earlier benchmark treatment, we adopt the same phenomenological prescription~\cite{YaserAyazi:2024dxk, Kamionkowski:1993fg,Caprini:2009yp}:
\begin{equation}
 \begin{aligned}
 \kappa&=\frac{0.715\alpha+\frac{4}{27}\sqrt{\frac{3\alpha}{2}}}
                    {1+0.715\alpha},\\
 \kappa_v&=\frac{\alpha}{0.73+0.083\sqrt{\alpha}+\alpha},
 \qquad \kappa_{\rm turb}=0.05\kappa_v.
 \end{aligned}
 \label{eq:efficiencies}
\end{equation}
Together with $v_w=1$, these choices define the GW estimate. They do not determine the wall-friction dynamics of the model. We illustrate the GW signal with the benchmark selected in the DM analysis. Its relevant particle and transition parameters are given in Table~\ref{table2}. The quoted temperatures satisfy $T_N/T_C\simeq0.458$, indicating that nucleation takes place after substantial cooling below the temperature of phase coexistence. The spectrum is evaluated using $T_N=249\,\mathrm{GeV}$,
$\alpha=0.16$, and $\beta/H_*=1207.65$. For these inputs, the efficiency factors are $\kappa\simeq0.1678$, $\kappa_v\simeq0.1733$, and $\kappa_{\rm turb}\simeq0.008666$. Our main results are shown in Fig.~\ref{GWs}, which displays the GW spectra predicted for the selected benchmark point. These spectra fall within the projected sensitivity ranges of the planned space-based detectors Big Bang Observer (BBO), Decihertz Interferometer Gravitational wave Observatory (DECIGO) and Ultimate-DECIGO (UDECIGO).
The collision and sound-wave peak frequencies are $f_{\rm coll}\simeq1.14\times10^{-2}\,\mathrm{Hz}$ and $f_{\rm sw}\simeq5.71\times10^{-2}\,\mathrm{Hz}$, respectively, while $f_{\rm turb}\simeq6.83\times10^{-2}\,\mathrm{Hz}$. Figure~\ref{fig:gw-bp1} shows the total spectrum and the three source terms. The sound waves provide the largest contribution near the total maximum, where $(\Omega_{\rm GW}h^2)_{\rm peak}\simeq1.25\times10^{-12}$. The collision term is small in this region because its amplitude contains $(H_*/\beta)^2$, while the sound-wave amplitude is proportional to $H_*/\beta$. The turbulent signal is further reduced by the small
conversion fraction and the factor $1+8\pi f/h_*$ in its spectral function. It is also useful to compare the transition temperature with the typical DM freeze-out scale. The conventional estimate $T_F\sim\max(M_V,M_{\psi_1},M_{\psi_2})/20$ gives approximately $161\,\mathrm{GeV}$, which is below $T_N$ for BP1. This ordering is
compatible with freeze-out in the broken phase, provided that the transition completes and the species remain in thermal equilibrium until their respective decoupling temperatures. The predicted maximum lies at about $0.057\,\mathrm{Hz}$, in a frequency region relevant to proposed space-based interferometers, including BBO
and DECIGO~\cite{Crowder:2005nr, Kudoh:2005as, Yagi:2011wg}. The spectrum can therefore provide a complementary test of the DM parameter region considered here. Its quantitative detectability depends on the detector sensitivity, observation time, foreground subtraction, and the lifetime of the GW sources. Within the adopted approximation,
BP1 demonstrates the connection between the thermal evolution of the common scalar sector and the phenomenology of all three DM components.

\section{Conclusion}\label{conclusion}
We have investigated a classically scale-invariant extension of the SM with a dark $U(1)_D$ gauge symmetry, a complex scalar, and two Dirac fermions. The dark sector contains three DM candidates: two fermions and a massive vector boson. Radiative symmetry breaking through the Gildener-Weinberg mechanism generates the scalar vacuum expectation values and the associated particle masses without introducing explicit mass terms at tree level. We have studied the relic abundances and direct detection signatures of the three components, examining how their masses, interactions, and relative abundances determine the phenomenology of the model.

Using the one-loop finite-temperature effective potential with daisy corrections, we have also investigated the thermal history for a representative benchmark. Within the approximations adopted, the potential exhibits a first-order transition along the mixed Higgs-dark scalar direction. This connects the DM scenario to a possible source of a stochastic gravitational-wave background. This GW falls within the observational window of experiments such as BBO, DECIGO, and UDECIGO, which may allow for its investigation in the future and could serve as a pointer toward new physics.

\section*{Acknowledgments}
R.R. is supported in part by the National Research Foundation of Korea (NRF) Grant No. RS-2026-25484206, and the BK-21 FOUR program through NRF.	
\newpage
\bibliography{References}
\bibliographystyle{JHEP}
\end{document}